# The Multiverse and the Origin of our Universe

Tom Gehrels

*Department of Planetary Sciences, University of Arizona*

(July 6, 2007)

This paper has a new model of the multiverse and of our universe's origin through strict application of a classical equation in atomic physics, and the following discoveries have resulted thus far. The multiverse is a hierarchy in the number of universes, increasing stepwise towards infinity. It is a trial-and-error evolutionary system, in which universes survive only near critical mass. That mass is actually a factor of 1.94 less than the critical mass, and this is found to be consistent with the baryon density inferred from nucleosynthesis in our universe; it is also precisely verified as a cosmological effect. That factor seems to have originated in the multiverse for causing intersecting expansions of its universes, such that *mixing* occurs of debris from aging galaxies (over proton-decaying time scales). It follows that there is an inter-universal medium (IUM), probably having the *demand* of new universes in balance with the *supply* of radiation and sub-atomic particles from the decaying galaxies. The mixing causes the universes to have the same quantum, relativity, gravity, and particle physics as our universe. It also follows from the mixing that the universes begin nearly isotropically uniform, with whatever small variations that individual galaxies may have. The making of a universe from the radiation and sub-atomic particles occurs through re-vitalizing the protons, and other particles as well, by gravitational energy obtained in accretion of the IUM. This process therefore begins wherever the IUM space density reaches proton density, $\sim 10^{18}$ kg m$^{-3}$. The process continues quietly as the sweeping-up and gravitational accretion continue, until the near-critical mass is reached. Some of the IUM debris must also be pervading our present universe, steadily or in partially accreted lumps. The model therefore predicts that the IUM's sub-atomic particles appear as our *dark matter*, and its radiation component as our *dark energy*, both near 0 K temperatures. The dark energy may cause expansion phenomena, in addition to the above non-flatness expansion, from an accretion lump that arrived at our universe at $t \sim 9 \times 10^9$ y. Finally, it does not seem feasible to model a beginning or an ending for such a vast and holistic multiverse. An extraneous discovery may be useful for particle physics, namely that if our universe were ever at Planck density, its size would have been that of the proton. That size is thereby derived as an equivalent radius of a sphere for the varying shapes of the proton, at 8.197 3725 x 10$^{-16}$ m, with its precision determined only by those of h, c, and H. The proton plays a prominent role in this model, such that its mass, H, appears to be a cosmological constant as well as h, c, and G, all four lasting at least over proton-decaying time scales. The Planck mass seems defined by its role in the multiverse, and the Planck constant appears to be h, rather than ℏ = h/2π.

## I. CHANDRASEKHAR'S DISCOVERIES

Subrahmanyan Chandrasekhar (1910-1995) derived a general equation of mass in terms of cosmological constants for his theory of the structure, composition and source of energy of stars [1]. He had developed that discipline with detailed laws such as of Stefan and Boltzmann, relating pressure and temperature at various depths inside the star, but for massive stars he derived,

$$M \approx (hc/G)^{1.5} H^{-2}, \tag{1}$$

showing that the ultimate control is by the physics represented by the Planck constant h,

the velocity of light c, gravitational constant G, and the mass of the proton, H, that is by quantum, relativity, gravity, and particle physics, summarized as atomic physics [1, 2]. He noted that, "The essential reason for the success of current theories of stellar structure based on atomic physics can be traced to this fact that $(hc/G)^{1.5}$ $H^{-2}$ is a mass of stellar order". A related theory applies to matter that is degenerate, in the sense of its high density being a function of pressure only, independent of temperature [3]. He discovered a generalization of Eq. (1),

$$M(\alpha) \approx (hc/G)^{\alpha} \; H^{1-2\alpha}, \qquad (2)$$

in which the exponent α identifies the type of object, for instance α = 1.75 for our galaxy and 2.00 for our universe, as well as the above α = 1.50 for a massive star. A simple equation could do that from physical constants determined for other applications in terrestrial laboratories! More professionally, he noted that this equation may indicate "deeper relations between atomic theory and cosmogony" [2]. He judged it important enough to publish it three times, in 1937, 1951, and 1989, but its time had not come for lack of firm data and accepted modeling in cosmology. I took his classes in the 1950s, kept in contact with him, and remembered the M(α) concept [4].

The Wilkinson Microwave Anisotropy Probe (WMAP, [5]) and other programs brought the essential data for comparison with M(α), while generally accepted theories have made the time further ripe for applying M(α) to other universes as well as to ours [6,7,8].

This paper has a preliminary version on universes in astro-ph/0701344, and is now aimed at the broader context of their multiverse. It begins with a simplification of M(α), and with its calibration for making precise predictions in Table 1 (Sec. II). Section III compares the predictions with observations for various objects in our universe. Section IV prepares for reaching out to other universes and Sec. V has confirmations to facilitate doing so. Section VI makes predictions (in terms of proposals for future work), and Sec. VII overviews the recycling of universes. Remarks about the conclusions are in Sec. VIII.

## II. THE POWER OF M(α)

Two limitations to membership of M(α) are defined first. Equation (3) is simplified, and it is calibrated by basing it on the proton and Planck masses. Table 1 is to serve as a mainstay for the discussion.

### A. Baryons and Original Objects Only

This paper deals with baryonic masses only, *i.e.* consisting of observable matter, which amounts to only 4% of the universe. The other components are 22% dark matter connected with galaxies, and 74% for some form of dark energy in the expanding inter-galactic space [5]. They will be identified in our universe as possible debris from decaying universes and be shown not to have constant percentages (Sec. VI E).

This paper has meaning only for mass-scaling that happens at the origins of cosmic masses. For instance in the case of stars, the usage of M(α) will be limited to matter consisting primarily of hydrogen and helium, rather than of the later compositions that have increased abundance of heavier elements in subsequent stars. In fact, the aging of the compositions points towards the universe's demise (Sec. VI D).

### B. Reliance on Planck and Proton Masses

First, a simplification of Eq. (2) is made by expressing the masses in the *universal* unit of the proton mass, such that H = 1, and

$$M(\alpha) = (hc/G)^{\alpha}, \qquad (3)$$

in proton masses. Even though the H-term is now gone, it is important to remember the presence of the proton mass in all applications (such as in Sec. V A).

The strengthening of their application is to calibrate Eq. (2) in Eq. (3) with the Planck mass, which is already used in cosmology (Sec. IV D). This establishes the equation on a powerful physical foundation because it has the involvement of Planck and proton masses in addition to h, c, and G. The Planck mass is computed with $(hc/G)^{0.50}$ and, together with the proton mass, they provide the firm calibration for Eq. (3) and its tabulation.

### C. Basic Tabulation

Table 1 is computed with Eq. (3), and it will be central to the discussions all the way to the end. The Table presents the data first in proton masses and then, in the third column, in solar masses (s.m.) or kilograms. The class of objects is next, and the last column gives the $\alpha$-values of the observations as they will be derived in Sec. III.

**Table 1. Predicted and Observed Values of M($\alpha$) and of $\alpha$**

| Exponent $\alpha$ | Predicted proton masses | units shown | Type of Object in M($\alpha$) | Observed $\alpha$ |
|---|---|---|---|---|
| 2.00 | 1.131 79 (35) x $10^{78}$ | 9.5172 x $10^{20}$ s. m. | Baryonic Universe | 1.998-2.008 |
| 1.75 | 1.981 73 (53) x $10^{68}$ | 1.6664 x $10^{11}$ s. m. | Young Galaxies | 1.72-1.77 |
| 1.50 | 3.469 96 (79) x $10^{58}$ | 29.179 s. m. | Early stars | 1.50-(1.53) |
| 0.50 | 3.261 68 (25) x $10^{19}$ | 5.455 55 (41) x $10^{-8}$ kg | Planck Mass | |
| 0.00 | 1 | 1.672 621 71 (29) x $10^{-27}$ kg | Proton | |

s. m. = solar masses; estimated standard deviations are in the brackets.

Below the center line are the Planck and proton masses obtained from CODATA combination of laboratory data [9]; h = 6.626 0693 (11) x $10^{-34}$ $m^2$ $s^{-1}$; c = 299 792 458 m $s^{-1}$ (in a vacuum, exact); G = 6.6742 (10) x $10^{-11}$ $m^3$ $kg^{-1}$ $s^{-2}$. The numbers in parentheses are estimated standard deviations; the relative standard uncertainty for G is 1.5 x $10^{-4}$.

### III. COMPARISON WITH OBSERVATIONS

This paper must first establish the capabilities of M($\alpha$) before considering its use outside of our universe, and the way to do that is to check how well it predicts the masses inside. A comparison is therefore made with observations of our primordial universe, galaxies and stars. Section III D has a search for other objects that might be included in $M(\alpha)$ and reports on a confirmation of the scaling law by other authors.

### A. Our Original Universe

There are two independent determinations of baryon density for our universe. To make the comparison of densities with total mass, one multiplies with a volume, of course, but which is the appropriate volume? There is the apparent volume of the presently observable universe, which takes all observational effects into account [10, 11]. However, present observational corrections are not applicable to the original theoretical value of the mass in Table 1. The comparison with the observed densities is therefore made without taking the observational complications of the expansion into account, and yet taking the expansion itself into account. That appears to be the volume of a spherical universe having radius $1.373 \times 10^{10}$ lightyears, consistent with the expansion-age determination for our universe of $1.373 \ (+.013, -017) \times 10^{10}$ years [5].

A baryon density is inferred from nucleosynthesis in our universe at time $t \sim 1$ min [12], the result ranges between 1.7 and $4.1 \times 10^{-28}$ kg m$^{-3}$, yielding between $9.27 \times 10^{77}$ and $2.24 \times 10^{78}$ proton masses, with $\alpha$ between 1.9979 and 2.0077. An observation of baryon density is made from spacecraft [5] at $4.19 \ (+.13, -.17) \times 10^{-28}$ kg m$^{-3}$, yielding $2.300 \ (+.096, -.126) \times 10^{78}$ proton masses, at $\alpha = 2.007 \ 89 \ (+.000 \ 46, -.000 \ 55)$. All differences between observed and theoretical $\alpha$ in Table 1 are primarily due to the uncertainties in the density determinations, while the effect of uncertainty in radius 1.373 is less, which confirms the comparison method. The lesser agreement with the results from Ref. (5) should be investigated with more refined methods [10, 11].

The present critical density (for a universe that is flat in the sense of being without gravitational collapse or excessive expansion) for baryonic matter is $4.0 \ (\pm.4) \times 10^{-28}$ kg m$^{-3}$ [Refs. 5, and 13 corrected for $H_0$ of Ref. 5], yielding a total mass of $2.2 \ (\pm.2) \times 10^{78}$ proton masses, at $\alpha = 2.008 \ (\pm.001)$. The mass in Table 1 is $1.94 \ (\pm.18)$ times smaller than the critical mass. Any of the constants would have to be a factor of 1.39 different (for instance, $G = 4.8 \times 10^{-11}$ m$^3$ kg$^{-1}$ s$^{-2}$), which is out of the question. Sections V C and VI will show that the factor of 1.94 is a real cosmological effect.

### B. Original Galaxies

Observations of the 21-cm hydrogen-line for a variety of spiral galaxies show $5 \ (\pm 4$ s.d.$) \times 10^{10}$ solar masses [14]. One should perhaps select the upper limit to allow for dissipation of energy and mass through collisions, but there is also accretion from dwarf galaxies [15]. Carr and Rees derive the upper limit of galaxies near $10^{12}$ solar masses [16]. Observation has also been made of young galaxies at great distance; this is for multiple galaxies occupying a single dark halo, and they total $10^{11}$-$10^{12}$ solar masses [17]. A dispersion in $\alpha$ of 1.72-1.77 is in the Table, representing the above range of $10^{10}$-$10^{12}$ solar masses. A question whether or not the $M(\alpha)$ model should include the galaxies at such relatively inexact observational status (compared to that of the universe), and not knowing the time at which to define original galaxies, is in Sec. IV C.

Critics have pointed out that the observed range for all galaxies is $10^7$-$10^{12}$ solar masses ($\alpha = 1.64$-$1.77$). However, that range includes subsequent development [15], while we are interested in original galaxies, the youngest. Stars show the same effect, of the masses for subsequent stars varying over three orders of magnitude, while the original masses having a small range.

### C. Primordial Stars

For checking the stars of $M(\alpha) = (hc/G)^{\alpha}$, physical laws of pressure and temperature within stellar interiors have established that the exponent $\alpha$ is exactly 1.50 [2, 16].

The selection of observations for the present comparison is from stars that have "early" spectral type O, and their values lie near 30 solar masses [18]. They are stable, without apparent variability in brightness, but they have short lives. They end in supernova explosions with two shockwaves, the first of radiation, followed by a slower one of matter. The latter delivers atomic nuclei to the interstellar medium, of atomic weight higher than those of the original hydrogen and helium.

Reports have appeared in the literature of much more massive stars, but they are either resolved as stellar clusters, or they consist of accreted masses, or of highly unstable and shedding mass, or they have much heavier than hydrogen-and-helium composition [19]. An extreme of 500 solar masses [20] is included in the Table as 1.53, but in parentheses to indicate doubts that the extreme fits the criteria of Sec. II A, or that is included in the derivation of Eq. (1).

Incidentally, "solar mass" in Table 1 merely indicates a unit of $1.9891 \times 10^{30}$ kg, rather than numbers of solar-type stars; the early types considered here are fewer and more massive.

### D. Search for Completion of $M(\alpha)$

In order to specify the $M(\alpha)$ model, an extensive search was made for other objects, provided they are primordial and constrained by Sec. II A. The search was not limited to specific values of $\alpha$ because $hc/G$ has the dimension of a mass at all values of $\alpha$.

The stars in open and globular clusters consist mostly of subsequent atomic nuclei, showing spectra later than those of type O.

The clusters of galaxies are included in the data for the second line of Table 1 as subsequent subdivisions. That is how they became grouped, for instance our Milky Way galaxy resides in a supercluster, subdivided further into the Virgo Cluster and the Local Group.

At $\alpha = 1.00$, planetesimals of rocks and soil at 1-km radius might appear to be original objects in the solar system [21, 22], but they do not have the type of material considered in Sec. II A.

No other members were found for $M(\alpha)$; Table 1 has all that are in our universe on the basis of membership being defined as in Sec. II A. Early-type stars are therefore the smallest primordial and therefore the most basic objects primarily of hydrogen and helium and stable enough for application of $M(\alpha)$.

Binggeli and Hascher [23] discovered a universal mass function following a power law of the form $M^{-2}$. This is a fundamental confirmation of $M(\alpha)$. Its values for original galaxies and primordial stars overlap with their Figures. $M(\alpha)$ extends the same power law into the multiverse, where it originated (Sec. VI G).

## IV. OTHER UNIVERSES

The topic of this paper is one of exploration and discoveries, which had begun of course with Chandrasekhar's (Sec. I). Numerically he noted that Eq. (1) yields $9.5 \times 10^{20}$ solar masses for the universe and, independently of that, $1.1 \times 10^{78}$ for its baryon number [1]. Equation (3) now invites consideration of α > 2.00 (Sec. IV A). Theories of inflation are already involved with other universes (Sec. IV B). Section IV C finds special importance of the Planck mass, and Sec, IV D does that for the proton in preparation for exploring the multiverse. The word 'multiverse' is defined as the ensemble of all universes.

### A. The Filling of Evolutionary Options

Application beyond α = 2.00 must take into account that M(α) is a mass at any value of α, as we did in Sec. III D. This type of exploration beyond our universe has no restriction such as by the finite velocity of light. In other words, we may do what Chandrasekhar did, going from where he knew the stars to be at α = 1.50, to α = 2.00 expecting a universe. We may expect (from knowing a universe to be at α = 2.00) a mass increment in M(α) at some value of α, and the indications are in Table 1 that it is at α = 2.50 (see Sec. IV C regarding α = 2.25).

Inorganic and organic evolutions have a predominant characteristic that when any option is open, it will be filled [24], and values of M(α) are indeed open for α > 2.00, all the way to infinity. In other words, nature and its predominant characteristic of evolution beckon us to consider an ensemble of $3.3 \times 10^{19}$ universes at α = 2.50 just as we considered $3.3 \times 10^{19}$ primordial stars for the universe [see Eq. (5), below].

This leads to the first discovery of the "M(α) Model". To begin with, M(α) tends to infinity, which seems appropriate for a multiverse. There is no stop, the possibility of α = 3.00 with $3.3 \times 10^{19}$ ensembles is as likely to have been filled as the previous step or any step at larger α. It is a hierarchy.

In summary of this Section, by taking the interval between Planck and proton masses into account as the basic step, Δα = 0.50, it is discovered that M(α) presents a series ranked with α for increasing numbers of universes.

### B. Other Multiverse Theories

The time is ripe for considering the multiverse because papers and books are appearing with various models for various reasons, not always scientific. As an example of the ones in science, there has been modeling since the late 1970s for a fast expansion of our universe when it supposedly occupied a small space at age ~$10^{-35}$ s, an "inflation". It is an intricate procedure with thorough processing towards the beginning of particle formation, removing a variety of uncertainties in the understanding of early stages for our universe. It particularly provides large-scale uniformity and isotropy to explain that they are observed in the present 3-K background radiation [5, 6].

Inflation theories have been developed into a well-established discipline, and WMAP observations provide confirmation [5,7]. The models have progressed to other universes, such that their masses are also modeled to have spawned from quantum fluctuations of a *space-time background*, "space-time" implying their unification in Einstein's theories [7,8]. The following Sections use the concept of spawning of universes from an inter-universal medium as an example of previous exploration beyond our universe.

### C. Definition and Role of the Planck Mass

A Planck Era has been described for a theoretical foundation of inflation theories [6]. The name of the era refers to a Planck-density phase in which a Planck mass can have most of its components interacting at velocity c (a Planck length in a Planck time) if they would reside within a cubic Planck length. The Planck density is $c^5/hG^2$ and for h, c, and G in Sec. II C, it is

$$8.2044 (18) \times 10^{95} \text{ kg m}^{-3}. \qquad (4)$$

The above concept has defined the Planck mass until now, but a newly realized definition for the Planck mass is its role in the mass scaling of the multiverse and its universes.

For such modeling of M($\alpha$) in consecutive masses, steps of $\Delta\alpha = 0.50$ should be taken, rather than of $\Delta\alpha = 0.25$, for the following reasons. First, nature seems to point to the step of 0.50 near the bottom of Table 1 with the Planck mass at $\Delta\alpha = 0.50$ separation from the proton mass. Second, even though the fit of galaxies to the prediction at $\alpha = 1.75$ in Sec. III B is good, relative to its large scaling interval of $\Delta\alpha = 0.25$, the fit is not as good as those for universe and stars. There is also a question of whether galaxies formed near maximum size, or that they accreted towards that size, and there may be a problem that births of stars and galaxies appear to be intertwined. It therefore seems prudent to leave the topic until additional observations and interpretations of early stars and galaxies are made. The delay causes no difference in the M($\alpha$) model, it is merely a temporary precaution, and the model may switch later to using $\Delta\alpha = 0.25$.

### D. The Special Role of the Proton

Andrei Sakharov computed the half-life of the proton to be finite, at more than $10^{50}$ years [25], and its increasingly difficult observational verification stands at $10^{35}$ years. Such half-lives seem appropriate for the active lifetime of galaxies, which are determined by aging through the transition from the original hydrogen and helium to heavier elements. The range in switching of time scales that is required to appreciate the action in a multiverse is not greater than what is already common in the sciences, from Planck times ($10^{-43}$ s) to the age of our universe ($10^{17}$ s).

There is however something mysterious about the proton that invites pursuit. While we expect quantum, relativity and gravity effects with h, c, and G during all the stages of our universe, including the early ones, the proton itself does not appear until age $\sim 10^{-6}$ s on the present clock (which is for the classical model of our sole universe). Particle physics is included, already since the 1930s, in a fundamental equation for our universe. The clock may need to be reset for an M($\alpha$) multiverse model.

### V. THE M($\alpha$) MULTIVERSE MODEL

The following four Sections provide an intensive and a broad vista for the M($\alpha$) model from detailed consideration of the proton. First, its radius is determined in three different methods in Sec. A, and the third brings the discovery that the near-critical mass criterion must have originated in the multiverse (Sec. B). Section C discusses how peculiar the proton results are, but Sec. D continues to confirm them by showing that the factor of 1.94 found in Sec. III A is a physical effect. The last Sec. E has an even more fundamental discovery, namely that our physics originated in the multiverse. These sections occur on the borderline towards future work because they contain new ideas that need verification.

### A. Determination of the Equivalent Proton Radius

The constant factor F between steps of $\Delta\alpha = 0.50$ in Eq. (3) is seen in the number of proton masses for the Planck mass, which is the same as the numbers of original stars in our universe and of other universes in the multiverse at $\alpha = 2.50$,

$$F = 3.261\ 68\ (25) \times 10^{19}. \tag{5}$$

F also allows the determination of a size for a universe in the Planck Era, at the Planck density of Eq. 4. (This is, however, not a suggestion that it actually happened, but merely a demonstration of the peculiar definition of the Planck density.) The ratio of the Planck and the universe's masses is the third power of F (in Table 1 one sees $\Delta\alpha = 1.50$ difference between $\alpha = 0.50$ and 2.00), but the third root of that is then taken in order to derive the length ratio from the volume ratio, coming back to F. Thus we obtain a size parameter for such a universe from F and the Planck length, $3.261\ 68 \times 10^{19} \times 4.051\ 32\ (30) \times 10^{-35} = 1.321\ 41 \times 10^{-15}$ m. That is however for a rib of the cube for Eq. (4), while we wish to obtain the radius for a spherical volume of the universe, if it were at Planck density, which is $R = 8.1974\ (8) \times 10^{-16}$ m.

A simpler verification, without demonstration of the cube in the definition of the Planck density, is to divide the mass of the universe by the Planck density of Eq. (4), and obtain that radius again.

A third and independent derivation is by realizing that the universe's mass, $(hc/G)^2\ H^{-3}$ kg [Eq. (2); H also in kg] divided by the Planck density, $c^5/hG^2$ kg m$^{-3}$ yields the volume of $h^3\ c^{-3}\ H^{-3}$ m$^3$, without the gravity term, G. After rib-radius conversion by a factor $(4\pi/3)^{-3}$, it follows that the radius of the universe, if it would ever have been at Planck density, would have been,

$$R = 8.197\ 3725\ (18) \times 10^{-16} \text{ m}. \tag{6}$$

The precision depends only on those of h and H, since c is exact and G is no longer involved.

### B. The Universes' Critical Mass originated in the Multiverse

We should try another choice for H (instead of the proton mass in Sec. II B), for instance the mass of the hydrogen atom as Chandrasekhar sometimes did. The result is $R = 8.192\ 9019 \times 10^{-16}$ m for the $^1$H atom. It is seen that for any value of H larger than that of the proton mass, R in Eq. (6) would be smaller, and vice versa (see Sec. V C).

It appears that evolution in the multiverse brought the exponent -3 to H for the above $h^3\ c^{-3}\ H^{-3}$ and in $(hc/G)^2\ H^{-3}$. Another way of expressing this is that the key trial-and-error evolution in the multiverse appears to be for $(hc/G)^2\ H^{-3} = 2.000$ and, together with $(hc/G)^2$, that is for $H^{-3}$. Paraphrasing Chandrasekhar in Sec. I, "The essential reason for the success of the M($\alpha$) model for a multiverse based on atomic physics can be traced to this fact that $(hc/G)^2\ H^{-3}$ is the mass of each of its universes".

### C. A Proton Mystery?

The proton mystery does not diminish, it deepens. The above paragraph appears to make H a cosmological constant as well as h, c, and G. We will see in VI F that the importance of the proton mass as a cosmological constant is even more enhanced by the decay and rebirth of protons, and thereby the decay and rebirth of universes.

However, smaller volumes (in m³) and radii (in m) for larger objects and vice versa is strange, as if Eq. (6) converges on the proton radius as some absolute law; Eq. (6) seems to express an absolute determination of the proton radius. Furthermore, the precision of R in Eq. (6) will increase evermore because it depends on those of h, c, and H – from terrestrial laboratory determinations – which are bound to be steadily improved with time. It is noted that c is also exact (Sec. II C). How else would we know that there is something meaningful in the apparent connection of the proton and the universe (and that at a density the universe probably never had)? There is something awesome about something that appears to be absolutely true, to have absolute truth. Here comes another confirmation that all this may indeed be true.

### D. Confirmation of the Non-Flatness of Universes

Equation (6) is precise enough to confirm a non-flat universe that has the expansion parameter of a factor 1.94 discovered in Sec. III A. First, there is observational confirmation in proton observations for Eq. (6). For a comparison with charge-radii observations, a straight average of the radius by various teams [26] gives 8.2 (±.3) x $10^{-16}$ m for six observations, of which there is one as far off as 6.4 x $10^{-16}$ m, while five are between 8.09 and 8.90 x $10^{-16}$ m. Two more observations yield 8.05 (±.11) and 8.62 (±.12) x $10^{-16}$ m [27]; their internal precisions are noted.

Second, we realize that the radius determination of Eq. (6) is with a new technique, but that this is of an *equivalent* proton radius, rather than the expression of the proton as a sphere. The proton has for a long time been considered a fuzzy sphere having radii between 6 and 10 x $10^{-16}$ m, but a better interpretation for the wide variation of precise measurements, is that its shape may be time-dependent perhaps due to internal quark motion [27, 28]. The word "equivalent" is then for a hypothetical spherical shape of the proton. The equivalent density of the proton, assuming uniformity, follows from Eq. (6) and from the mass of the proton in Table 1,

$$\text{proton density} = 7.249\ 1170\ (50) \times 10^{17}\ \text{kg m}^{-3}. \qquad (7)$$

Finally, we note that the proton results confirm the finite mass of the universe at 1.131 79 x $10^{78}$ protons precisely, and that mass lies inside the range of the inference from nucleosynthesis in Sec. III A. On the other hand, the critical mass of 2.2 (±.2) x $10^{78}$ protons yields a proton radius of 1.02 (±.03) x $10^{-15}$ m, which seems unlikely when compared to the above observations, and is out of the question when compared to Eq. (6) and its precision. This determination of the proton radius yields a magnification of the factor of 1.94 discovery in Sec. III A; the radius relation is steep so that even a small deviation, as in α = 2.008, for the mass of the universe occurs outside the ranges for observations and derivation of R. The cosmological meaning of the 1.94 effect is in Sec. VI E.

### E. Our Physics originated in the Multiverse

Equation (3) expresses a mass-scaling law, but Table 1 also shows some order of time and evolution, from top to bottom. A curious discovery occurs when considering α = 0.50 and 0.00 in the Table, with their Planck and proton masses; Chandrasekhar, for example, did not consider the Planck and proton masses, he did not have Table 1. It is clear now that these physical parameters have come to our universe from before Table 1, from the outside. By considering also the parameters of Eq. (2), it is shown that these are quantum, relativity, gravity, and particle physics, as Chandrasekhar said, atomic physics.

Two conclusions are unavoidable, namely that these physics are also the ones of the inter-universal medium from which our universe spawned, and if it happened this way for our universe, it must be the way for all universes of $M(\alpha)$. If indeed all universes came about from the mixing of debris from many galaxies (Sec. VI E), it confirms that the physics and near-critical mass are the same for every one of the universes in the $M(\alpha)$ multiverse. In other words, a multiverse described by $M(\alpha)$ must have its physics and mass scaling. Further discussion occurs in Sec. VI G.

## VI. PREDICTIONS AND FUTURE WORK

The following seven Sections mention predictions, but because they are all so new they need scrutiny and follow-up by experts, and they are therefore written as suggestions for future investigations.

### A. Observational and Theoretical Work

A practical application of the $M(\alpha)$ model is to stimulate observations of early galaxies with powerful instrumentation in order to remove the uncertainties in Secs. III B and IV C. The same applies to original stars in order to settle the question of their stability and $M(\alpha)$ membership (Sec. III C). The broader quest is for observations of the other universes, remotely, indirectly, or otherwise; their dark matter and energy may already be observable (Sec. VI E). Could the WMAP be used to check on aspects of the $M(\alpha)$ model? As for theoretical applications, the model needs scrutiny and derivation of detailed expressions and their solutions.

The equivalent radius of the proton, with high precision as in Eq. (6), may be of interest to particle physicists [27, 28]. It may help determine the shape(s) of the proton through combination with observations of proton size made at various times, and perhaps bring information on quarks.

### B. The Hierarchy as a Quantization

The numbering of the alphas in $M(\alpha)$ seems an anthropic experience, for we think of our universe having $\alpha = 2.00$, and another society in our galaxy, or in another galaxy or universe, will do the same. Our universe and theirs are then imagined by us and by them as a member of an assembly of $3.3 \times 10^{19}$ universes at $\alpha = 2.50$, which is imagined as a member of $3.3 \times 10^{19}$ assemblies at $\alpha = 3.00$, and so forth [see Eq. (5)].

In such imagination, it also seems logical to consider that we see more universes at greater distance, and that still would be an anthropic experience.

$M(\alpha)$ is a quantization concept, illustrated and made better understandable in fact by these two experiences.

### C. No Beginning – No End

The argument regarding "no beginning" between Ref. (8) and Ref. (7) may be re-visited by taking the physics of h, c, G and H into account. In addition to the power of such unified physics, one should consider the enormous scale of the $M(\alpha)$ multiverse having an infinite number of universes, and an infinity of encounters, collisions, and interactions as well. The mechanism for the latter three is illustrated in Sec. VI E. Here, the origin of all physics and characteristics of our world is considered, namely that the three are the mainstay of evolution in trial-and-error search for possibilities of survival, a primary topic in studies of evolution [24].

The M($\alpha$) model shows that the multiverse is without a beginning or end because of the above infinities. There appears to be no scientific way to start such a vast and holistic multiverse all at once, or to develop it incrementally, or to stop it. If so, time has no beginning or ending either, because of the multiverse' all-encompassing influence and magnitude in continuing action and movement. The multiverse is alive with motion and interconnections. These will be challenging topics for scientific investigation.

### D. Accretion towards making Universes

The M($\alpha$) model considers the *demand* of newly spawned universes to be in balance with the *supply* of mass and energy from decaying universes. A lead for this topic may be Sakharov's half-life limit for the proton (Sec. IV D), which implies decay and perhaps loosening of atomic bonds. If the latter is verified, the effect will be faster for the larger nuclei due to their weaker bonds. On the other hand, elementary particles may not decay, or if they do, it would occur on an even longer time scale than that of the proton, for which Sakharov wrote already "very large (more than $10^{50}$ years)". The surviving elementary particles end in the IUM, individually, in grains or clumps, or as nuclei of white dwarfs for example, or perhaps even as whole decayed galaxies. They will have densities near the $10^{18}$ kg m$^{-3}$ of Eq. (7), and that may be the general density of the IUM material, at low temperatures. The basic interaction mechanism is the mixing effect of Sec. VI E. The grains and lumps or clumps grow by sweeping material together, and eventually continuing the growth gravitationally, as happens for interplanetary and interstellar matter.

Great promise appears in studies of mechanisms that are likely to happen because intergalactic and inter-universal space must be receiving radiation, energy, neutrinos, cosmic rays, and matter debris from aging galaxies and clusters of galaxies. It needs to be investigated how much leakage there is away from the galaxies' gravitation and how observable the effects are. In the meantime, we ought to use the name of *inter-universal medium* (IUM, from radiation and sub-atomic particles) instead of, or together with *space-time background*, from quantum fluctuations (Sec. IV B)

Sakharov's paper may need to be refined, and the topic of proton decay revisited, to verify his large value for the proton's half life. Incidentally, reports of cosmological variation of the fine-structure constant [29] seem not in conflict with the M($\alpha$) model because that variation is likely to be also an aging effect, if real at all (between -2.5 and +1.2 x $10^{-16}$), to be considered together with that of the proton decay.

Observations of the intergalactic medium (IGM) within and outside of galaxies and clusters of galaxies may be informative. The question is if there is indeed a closed loop of losses due to the spawning of universes matched by gains from debris of galaxies that are aging due to their transition of hydrogen to heavier elements. The accretion may be compared to that in the interstellar medium (ISM) also regarding supernovae bow shocks, but having differences in particle density to begin with, in composition of the components, and in energies of the collisions through the mixing effect. A related question is to what extent the multiverse acts like a closed box, with all its conservation laws. Entropy needs to be included - perhaps the multiverse ages too, and its debris makes another multiverse, and so forth.

## E. Our Universe's Dark Matter and Energy

Sections III A and V D conclude that the factor 1.94 is a real cosmological effect, the "1.94 effect". The cosmological meaning is that each universe with its galaxies thereby expands through the others. An example of interaction by expansion is seen in two dimensions from above a quiet pond on which various raindrops fall. The rings expand and travel through each other, and the result in three dimensions must be a thorough mixing of the debris from old galaxies. This is referred to as the "mixing effect". It brings interaction and endless opportunities of encounters by and for all universes.

The factor of 1.94 may be sustained in trial-and-error of continuing evolution in the multiverse, because perfect flatness would not yield interaction of universes for the system to survive. In first order, this explanation assumes that there is no expansion of the inter-universal space as there is for intergalactic space, but if there were, the evolution within the multiverse would have included that in the factor.

Our universe is of course immersed in the inter-universal medium (IUM), as well as all other universes are. The question arises whether some of the IUM's sub-atomic particles and radiation can be observed in our universe too. In principle, the debris from decaying galaxies ripples in all directions on the expansion of intergalactic space for each universe, overlapping in the above mixing effect, reaching everywhere, and that includes our universe. There are some similarities with the ISM, which is supplied by the usual two waves from supernovae, namely the matter wave travelling slowly, preceded by the energy wave travelling at the speed of light. The IUM would on average be rather uniform, but it may be locally uneven in space density as are the molecular clouds of the ISM, perhaps occurring in clumps. So it is a matter of luck how much our universe has received until now.

The clouds or clumps are gravitationally captured by galaxies, including our own, and that then may be our *dark matter* and *dark energy*. Dark matter was discovered by J.C. Kapteyn in 1905, studied by J.H. Jeans, J.H. Oort [30], by F. Zwicky [31], and by others up to the present time. It used to be called *missing mass*, and understood as aging debris of our own galaxy such as in exhausted white dwarfs, but that did not amount to enough, and this problem may now be solved. Its density in our galactic space seems greater than that of our baryon density, so our luck is to have caught a hefty clump. The $M(\alpha)$ model predicts that it consists of whatever is left after dissolution of atomic nuclei. The observational problem will be to distinguish it from the hydrogen that is already known to be there. Direct collection perhaps? The subatomic particles might be collected and analyzed by spacecraft, as has been done for grains within the solar system. Pioneers 10 and 11 and the Voyagers are examples of spacecraft that have left the domain of the solar system to enter galactic space.

The dark energy might consist of photons from old but originally bright sources such as active galactic nuclei; their aging is in terms of shifting to temperatures close to absolute zero. This debris might now be present everywhere in the expanding intergalactic space of our universe. The question is whether this can be compared to what is called the *dark energy* found by WMAP and other observatories. Some theorists have appropriated it as the *cosmological constant* as another name for it, implying that it drives expansion phenomena. The $M(\alpha)$ concept of debris may be sufficient for such drive because its energy may cause expansion. A hefty clump may have accelerated the intergalactic expansion as has been observed at our universe's age of $t \sim 9 \times 10^9$ y; perhaps it is the same hefty clump mentioned in the previous paragraph. The percentages of 22% dark matter and 74% dark energy mentioned in Sec. II A appear to be not constant; they probably are variable with time and place in the accreting IUM. However, the starting mass of a universe appears to be constant, at $\alpha = 2.000$, although in further decimal places there may be variation as far as we know at this time; the exponent $\alpha = 1.5$ is exact (Sec. III C).

### F. The Beginning of our Universe

It appears necessary for the accretion in the IUM to reach $\alpha = 2.000$ in order to make a universe. In stellar studies that is an impossibly large mass [2]. $10^{21}$ solar masses is enormous - could such a large mass exist? The total-mass-plus-energy may be greater by a factor 25, envisioned as the baryonic mass, dark mass, and dark energy, all coming through that stage.

Here, however, the deciding condition here is that the material is degenerate, in the sense of its high density being a function of pressure only, independent of temperature. Chandrasekhar's modeling for degenerate matter may of interest even when it may not develop into a stellar object [3]. The accreted mass may start acting towards making a universe with a very different proposition than has been modeled for our sole universe, namely at the epoch when the space density of the gravitational accretion reaches the $\sim 10^{18}$ kg m$^{-3}$ level of Eq. (7), anywhere within its universe cloud of mass and energy. The re-fabrication to protons etc. might begin at that epoch near the gravitational center of the cloud and proceed steadily with restoring and revitalizing of protons and other particles. Expansion is probably associated with the processing, in conflict with the gravitational accretion. That conflict may control the process, in some analogy with the thermostat effect of star formation from an interstellar cloud.

All this needs to be modeled by experts. It will be rewarding to establish just how, and how long before the time of proton (re)formation, the understanding of our universe will be back on its present track. In other words, at what time on our present clock (for our sole universe) does the beginning of the M($\alpha$) universe occur? For comparison, the time of proton formation is t ~ $10^{-6}$ s on that clock.

The M($\alpha$) model shows that the finite baryonic mass, to be taken from the original cloud of dark mass and energy, is $\alpha = 2.00$. If in the subsequent modeling again the annihilation of particles and anti-particles occurs, there may be a problem arising from taking two data sets into account. On the one hand, the epoch at the end of the inter-universal accretion at $\alpha = 2.00$ seems indicated. On the other hand, the agreement in Sec. III A with Ref. (12), indicates that at a time after nucleosynthesis, when our universe had age near one minute, the baryonic mass would still be $\alpha = 2.00$. This problem should be interesting for particle physicists.

### G. Evolution originated and is sustained in the Multiverse

Astrophysicist Fred Hoyle pointed out decades ago that the probability is extremely low to have the fine-tuning that is observed for the nuclear transitions within stars, which ultimately may produce the extreme complexities of life. If the physical constants of the elements would have been even slightly different, the selections and combinations would not have occurred. This problem seems now solved because the continuing evolution within the multiverse keeps only finely tuned universes to begin with, of critical mass and of unified quantum, relativity, gravity and particle physics.

Similarly, the uniformity problem (for our own universe, if it were the only one) seems to have been solved because the IUM material is homogenized throughout its history through the accretion from many galaxies in the mixing effect. This even allows for the galaxy non-uniformities observed by WMAP and others, because the accretion was from the same type of individual galaxy variation to begin with. Other classical problems are more relaxed as well, for instance that the flatness parameter in the beginning had to have been equal to 1.0 within one part in a huge number, because now our universe did not begin in a small volume and its mass was near critical to begin with.

Careful analysis needs to be made of the accretion in the IUM, of a body reaching $\alpha = 2.00$, of the sweeping effects by gravitational cross-sections for such growth, and how that body would come to re-constitution of protons and other particles, undoing their decay in aging universes. Larger bodies than for $\alpha = 2.00$ can probably not accrete, so the case of universes more massive than the critical mass may not occur; if they did, gravitational collapse of its subsequent universe would quickly eliminate that deviation. Smaller accreted bodies and their universes, of less than critical mass, result in rapid demise through fast expansion, and participating in the general accretion again. Trial-and-error evolution thereby occurs, towards surviving universes all having the same near-critical mass; the survival of the system apparently occurred for an increased efficiency of the process at an expansion and interaction factor of 1.94 less than the critical mass,. Our basics, h, c, G. H physics, and all their derived laws, are also tested and reconfirmed in the evolutionary trial-and-error search for survival [24]. Everything originated in the survival evolution of the multiverse, including evolution itself.

The ultimate challenge is the explanation of the physical mechanisms of evolution and mass scaling. Quantization is a basic feature of our world, or it would be an amorphous brew. Binggeli and Hascher's power of 2 is important here (Sec. III D); their observations show that later developments run along that power law, and that would also be the case in the multiverse for accreted universes of less than critical mass. The fundamental question is of course why the formation of a universe should begin when the accumulation of baryonic mass reaches $\alpha = 2.00$. However, questions regarding the maximum mass of galaxies and stars have already been studied [16]. Numerical studies of these topics are needed, and there are so many that it looks like a new discipline for studying the multiverse could emerge.

## VII. RECYCLING OF UNIVERSES

The recycling of universes could be proposed independently of this paper, based solely on the assumption of degenerate mass, radiation, and perhaps whole galaxies decaying into the IUM. The galaxies may leak their radiation and aging matter beforehand already, and that effect might be observable in the intergalactic medium, as observations in the interstellar medium were developed during recent decades. If old universes indeed seed the IUM with radiation and sub-atomic particles, the new universes may begin long after t = 0 (on the present clock) because the sub-atomic particles have densities near $10^{18}$ kg m$^{-3}$ and not extremely high temperatures due to gravitational accretion.

Once the concept of recycling is accepted, the M($\alpha$) model would give detailed information about the multiverse. Previous cosmological models (for our universe being the only one) did not have much to say about what occurred before t = 0, or about the origin of our basics and physics, or about the peculiar importance of the proton, or about the cause of the sudden appearance of expansion acceleration at t ~ 9 x 10$^9$ y. The M($\alpha$) model can help with these points.

However, even without the M($\alpha$) model, the multiverse has known finite mass for its universes, because they can survive only near critical mass. It is, in fact, a requirement for any multiverse model that its universes have near-critical mass. Our universe has the physics we know so well, and the other universes must in any model have obtained the same physics if they were spawned from the same IUM background.

## VIII. CONCLUSIONS

This paper began merely with the fascination with Chandrasekhar's discoveries, and this lead to additional discoveries. The growth of the M($\alpha$) model can be seen in the order of this paper and in comparison with astro-ph/0701344. During the years of development for the M($\alpha$) model, new observations have invariably been close to the predictions, they brought deeper insight and further progress, and this process has not stopped. These are indications of coherence and perhaps even truth of the model, as are its common sense, internal consistency, and beauty [32]. Equation (3) is not a simple expression, it is tempered by atomic physics, and by Planck and proton masses. The M($\alpha$) conclusions follow directly from Eq. (3), like it or not, without any reliance on extraneous dimensions or model parameters selected towards a goal.

It seems an anthropic concept that our universe is the only one; the argument of it being infinite is still heard to imply that all of space is occupied by our universe. Instead, the *expansion* is towards infinity, for each universe, thereby bringing essential and energetic interaction of radiation and matter from and to all universes.

The M($\alpha$) multiverse seems to prove the constancy of the cosmological constants h, c, G, and H. Or, are they also rejuvenated, perhaps indirectly, in the multiverse' evolutionary process? The present treatment and its comparisons with observations indicate that the Planck constant is h, not $\hbar = h/2\pi$. The usage of $\hbar$ appears to have been initiated by Paul Dirac (1902-1984).

A way to visualize the multiverse may be as a giant tree with many small leaves, seen from anywhere so deeply inside the tree that its outer edge is not seen, each leaf representing a universe. The stems and branches of the tree are not continuous, but seen as isolated clumps in various stages of accretion, while a windy dusty desert day may show the general background of the IUM.

Chandrasekhar was adamant with his students that they understood and would apply the Reciprocity Principle [4]. He taught them to inspect data sets for reciprocities, and one he would have enjoyed is seen here. The paper proceeded upwards in Table 1 beyond our own universe, but reciprocity is now to consider the downwards direction. We found the multiverse from characteristics of our own universe, but that multiverse actually produced our universe to begin with. It follows that the multiverse provided our universe with its energy, mass, and physics. The application of the RP is that our observations of the latter three apply for the multiverse and all its surviving universes. The RP also provides an ultimate confirmation of the M($\alpha$) model.

A worldview emerges whereby quantization brings a self-tuned multiverse of surviving universes, which have the capability of life, even though no anthropic concept is used in this paper. The mass scaling may be useful in evolution and philosophy as well as in cosmogony and atomic theory.

**ACKNOWLEDGMENTS**

I thank Anthony Aguirre, Neil Gehrels, Piet Hut, Vincent Icke, Carl Koppeschaar, Kevin Moynahan, Masami Ouchi, Kevin Prahar, François Schweizer, Fred Spier, Erick Weinberg, and an anonymous referee, for their valuable comments.

[1] S. Chandrasekhar, Nature **139**, 757 (1937); also in Selected Papers, Vol. I. Stellar Structure and Stellar Atmospheres (Univ. Chicago Press, Chicago, Illinois,304, 1989).
[2] S. Chandrasekhar, in "Astrophysics, a Topical Symposium", J. A. Hynek, ed. (McGraw-Hill, New York, N.Y. 598 (1951).
[3] S. Chandrasekhar, Mon. Not. Roy. Astron. Soc*.* **91**, 456 (1931).
[4] T. Gehrels, On the Glassy Sea, in search of a worldview. (BookSurge, Charleston, South Carolina, 2007), original published by Am. Inst. Phys., New York, NY, (1988).
[5] D. N. Spergel *et al.,* Astrophys. J. Suppl. **170, No.2**, 377 (2007).
[6] A. H. Guth, The Inflationary Universe (Addison Wesley, New York, NY, 1997).
[7] A. H. Guth, in "Measuring and Modeling the Universe"*,* W. L. Freedman, ed. (Cambridge Univ. Press, Cambridge U.K., 2004).
[8] A. Aguirre & S. Gratton, Phys. Rev. D **67,** 083515 (2003), gr-qc/0301042.
[9] P. J. Mohr & B. N. Taylor, Rev. Mod. Phys. **77**, 1 (2005).
[10] C. H. Lineweaver & T. M. Davis, Sc. American **292, No. 3**, 36 (2005).
[11] A. Riazuelo, I. P. Uzan, R. Leboucq & J. Weeks, Phys. Rev. D **69**, 103514, Sec. V A (2004).
[12] C. J. Copi, D. N. Schramm & M. S. Turner, Science **267**, 192 (1995).
[13] S. Dodelson, Modern Cosmology (Academic Press, Burlington, Massachusetts, 2003)
[14] A. N. Cox, Ed., Allen's Astrophysical Quantities*,* 4[th] ed. (Springer Verlag, New York, NY, 579, 2000).
[15] F. Schweizer, F. 2000, Phil. Trans. R. Soc. London A **358,** 2063 (2000).
[16] B. J. Carr & M. J. Rees, M. J., Nature **278,** 605 (1979).
[17] M. Ouchi *et al*., Astrophys. J. **635,** L117 (2005).
[18] S. W. Stahler, F. Palla & P. T. P. Ho, in *"*Protostars and Planets IV", V. Mannings, A. P. Boss & S. S. Russell, eds. (Univ. Ariz. Press, Tucson AZ, 2000), 327.
[19] I. A. Bonnell, S. G. Vine & M. R. Bate, Mon. Not. R. Astron. Soc. **349**, 735 (2004).
[20] V. Bromm, V., Sky and Tel*.* **111, No. 5,** 30 (2006).
[21] H. Alfvén & G. Arrhenius, Evolution of the Solar System. (NASA SP-345, Washington, D.C., 1976).
[22] J. Kleczek. The universe. (Reidel Publ., Dordrecht, Netherlands, 1976).
[23] B. Binggeli & T. Hascher, astro-ph/07051599, accepted by Publs. Astron. Soc. Pacific.
[24] T. Gehrels, Survival through Evolution, from Multiverse to Modern Society (BookSurge, Charleston, South Carolina, 2007).
[25] A. D. Sakharov, Zh. Eksp. Teor. Fiz. **49**, 345 (1965); J.Exper.Theor.Phys. **22**, 241 (1966).
[26] S. G. Karshenboim (2000), http://arxiv.org/PS_cache/hep-ph/pdf/0008/0008137.pdf.
[27] D. J. Berkeland, E. A. Hinds & M. G. Boshier, Phys. Rev. Letters **75**, 2470 (1995).
[28] P. F. Schewe & B. Stein, Physics News Update, Phys. Today **242**, Sept. 25 (1995).
[29] H. Chand, R. Srianand, P. Petitjean & B. Aracil. Astron. Astrophys. **430**, 47 (2005).
[30] J. H. Oort, in "Galactic Structure", A. Blaauw & M. Schmidt, eds. (Univ. Chicago Press, Chicago, Illinois, 1965), and references therein.
[31] F. Zwicky, Morphological Astronomy (Springer Verlag, Berlin, 1957).
[32] S. Chandrasekhar, Truth and Beauty: Aesthetics and Motivations in Science (U. Chicago Press, Chicago, Illinois, 1987).